\documentclass[12pt,preprint]{aastex}
\usepackage{emulateapj5,epsf}

\usepackage{lscape}

\shorttitle{
On the origin of FRI/FRI\hspace{-.1em}I dichotomy
}
\shortauthors{Kawakatu, Kino and Nagai}

\begin{document}

\title{
On the origin of Fanaroff-Riley classification of radio galaxies: 
Deceleration of supersonic radio lobes
}


\author{Nozomu Kawakatu\altaffilmark{1}}
\affil{National Astronomical Observatory of Japan, 2-21-1 Osawa, 
Mitaka, Tokyo 181-8588, Japan}

\author{Motoki Kino}
\affil{National Astronomical Observatory of Japan, 2-21-1, 
Osawa, Mitaka, Tokyo 181-8588, Japan}

\author{Hiroshi Nagai}
\affil{National Astronomical Observatory of Japan, 2-21-1, 
Osawa, Mitaka, Tokyo 181-8588, Japan}


\altaffiltext{1}{kawakatu@th.nao.ac.jp}

\begin{abstract}
We argue that the origin of `FRI/FRI\hspace{-.1em}I dichotomy' --- 
the division between Fanaroff-Riley class I (FRI) with subsonic lobes 
and class I\hspace{-.1em}I (FRI\hspace{-.1em}I) radio sources with 
supersonic lobes is sharp in the radio-optical luminosity plane (Owen-White diagram) --- can be explained by the deceleration of advancing radio lobes. 
The deceleration is caused by the growth of the effective cross-sectional 
area of radio lobes. 
We derive the condition in which an initially supersonic lobe turns into 
a subsonic lobe, combining the ram-pressure equilibrium between 
the hot spots and the ambient medium with the relation between 
``the hot spot radius" and ``the linear size of radio sources" 
obtained from the radio observations. 
We find that the dividing line between the supersonic lobes and subsonic ones 
is determined by the ratio of the jet power $L_{\rm j}$ to the number density 
of the ambient matter at the core radius of the host galaxy 
$\bar{n}_{\rm a}$. It is also found that there exists the maximal 
ratio of $(L_{\rm j}/\bar{n}_{\rm a})$ and its value resides in $(L_{\rm j}/\bar{n}_{\rm a})_{\rm max}\approx 10^{44-47}\,{\rm erg}\, {\rm s}^{-1}\,{\rm cm}^{3}$, 
taking account of considerable uncertainties. 
This suggests that the maximal value $(L_{\rm j}/\bar{n}_{\rm a})_{\rm max}$ 
separates between FRIs and FRI\hspace{-.1em}Is. 

\end{abstract}

\section{Introduction} 
Fanaroff \& Riley (1974) discovered that the radio galaxies whose linear size 
$l$ is $l \geq 10\,{\rm kpc}$ exhibit a change in morphology from edge-darkened to edge-brightened at a monochromatic power $\sim 10^{24.5}$ W Hz$^{-1}$ at a 
rest frame frequency of 1.4 GHz. 
The Fanaroff-Riley type I radio galaxies (FRIs) have the edge-darkened 
morphology and the subsonic advance speed of their lobes (hereafter 
subsonic lobes), while the Fanaroff-Riley type I\hspace{-.1em}I radio 
galaxies (FRI\hspace{-.1em}Is) possess the edge-brightened morphology and 
the supersonic advance speed of their lobes (hereafter supersonic lobes). 
Owen \& White (1991) and Ledlow \& Owen (1996) found a striking separation 
between the FRIs and FRI\hspace{-.1em}Is in the radio-optical luminosity plane. At given optical luminosity of host galaxies, FRI\hspace{-.1em}Is are located 
on the side of brighter radio luminosity, while FRIs tend to fall on the side 
of fainter radio luminosity. 

The origin of the `FRI/FRI\hspace{-.1em}I dichotomy', which is an outstanding 
issue in the astrophysics of extragalactic radio sources, has long been 
debated. Two promising interpretations for the dichotomy 
have been proposed; 
One is that the dividing line can be understood by ``the intrinsic differences 
of the jet's kinetic power" between FRIs and FRI\hspace{-.1em}Is (e.g., Meier et al. 1997; Ghisellini \& Celotti 2001; Marchesini, Celotti, \& Ferrarese 2004). 
The other is that ``the deceleration process of the jets" can determine 
the dividing line, supposing that FRIs are initially supersonic 
(e.g., Falle 1991; Bicknell 1994; 
Kaiser \& Alexander 1997; Nakamura et al. 2008). 
The deceleration scenario are also supported by the following two observational facts; One is the observations of relativistic sub-kpc jets in nearby FRIs 
such as M87 (Biretta et al. 1995) and relativistic sub-pc ones in BL Lac 
objects (e.g., Gabuzda et al. 1994), whose parent populations are FRIs in 
AGN unified model (e.g., Urry \& Padovani 1995). On the other hand, the 
subsonic lobes at $l > 10\,{\rm kpc}$ are observed in FRIs. 
The second is the existence of a new class of double 
radio sources in which the two lobes have clearly different FR morphologies 
at both sides called `HYbrid MOrphology Radio Sources (HYMORS)' (e.g., 
Gopal-Krishna \& Wiita 2000). 
However, the deceleration process have been unclear.

In this paper, we especially focus on the deceleration process of lobes 
to explore the origin of `FRI/FRI\hspace{-.1em}I dichotomy'. 
{\it How does the supersonic flow decelerate in the first few kpc ?} 
As a plausible deceleration process, the entrainment through mixing in a turbulent layer between the jets and the ambient medium has been considered 
(e.g., De Young 1993a, b; Bicknell 1994; Perucho \& Marti 2007; Rossi et al. 2008). De Young (1993a, b) suggested that the deceleration of the advancing lobes due to the entrainment is important for the special cases, i.e., both relatively low kinetic power of jets and sufficiently dense ambient 
medium. In this paper, we elucidate another process of deceleration of radio lobe's advance speed. As a possible deceleration process, Cioffi \& Blondin (1992) showed the importance of the jet's head growth by using hydrodynamic simulations. Interestingly, recent observation has suggested that the cross-sectional area of reverse shocked region of jets (hot spots) grows faster in the host galaxy 
than outside the galaxy. By comparing the dynamical model of radio lobes and hot spots including their head growth, Kawakatu, Nagai \& Kino (2008; hereafter KNK08) suggested that this trend that hot spot radius changes with distance 
can be explained by the strong deceleration of radio lobes.

The goal of this paper is to elucidate the condition in which initially 
supersonic lobes become subsonic lobes, which can be closely linked to 
the origin of `FRI/FRI\hspace{-.1em}I dichotomy', {\it by considering that the 
deceleration of radio lobes and hot spots controlled by the growth of 
radio lobes' head}. 
In $\S2$, we provide a simple treatment for the dynamical evolution of radio lobes, in order to derive the dividing line between FRIs and FRI\hspace{-.1em}Is. 
In $\S3$, we show the ratio of the jet power and the constant number density 
of the ambient matter is maximal at the core radius. 
Then, it is newly predicted that the division between FRIs and 
FRI\hspace{-.1em}Is is determined by the maximal ratio of the jet power $L_{\rm j}$ to the number density of the ambient matter at the core radius of the host galaxy $\bar{n}_{\rm a}$. In $\S 4$, we suggest an evolutionary sequence of variously sized radio galaxies based on our findings. 
Summary is given in $\S 5$.

\section{Model and Observation}
In order to obtain the requirement for an initially supersonic lobe to turn 
into a subsonic lobe, we combine the ram-pressure equilibrium 
between the hot spots and ambient medium with the observed relation 
between ``the hot spot radius" and ``the linear size of radio sources" (Figure 1 in KNK08). 

We consider here a pair of relativistic jets 
propagating in an ambient medium $\rho_{\rm a}(l_{\rm h})$ 
where $l_{\rm h}$ is the distance from the jet apex. 
The equation of motion along the jet axis, i.e., 
the momentum flux of a relativistic jet is balanced to the ram 
pressure of the ambient medium spread over the effective cross-sectional 
area of the cocoon head, $A_{\rm h}(l_{\rm h})$, 
\begin{eqnarray}
L_{\rm j}/c&=&\rho_{\rm a}(l_{\rm h})v_{\rm HS}^{2}(l_{\rm h})A_{\rm h}
(l_{\rm h}),
\end{eqnarray}
where $L_{\rm j}$ and $v_{\rm HS}(l_{\rm h})$ are the total kinetic energy of 
jets and the hot spot velocity, respectively. 
Here, we assume that $L_{\rm j}$ is constant in time during the activity 
period $t\approx 10^{6-8}$ yr, which is a typical age of FRIs and 
FRI\hspace{-.1em}Is (e.g., Carilli et al. 1991; Parma et al. 1999; 
O'Dea et al. 2009; Machalski et al. 2009). 
Hereafter, we do not consider the entrainment effect of the ambient medium. 
This is justified for higher $L_{\rm j}$ (e.g., Scheck et al. 2002) 
but we will discuss this later on. 
At the hot spots, the flow of shocked matter is spread out by the oblique shocks (e.g., Lind et al. 1989) and the vortex via shocks (e.g., Smith et al. 1985). 
Thus, the effective ``working surface'' for the advancing jet is larger than 
the cross-sectional area of hot spots (Begelman \& Cioffi 1989; Kino \& Kawakatu 2005:hereafter KK05; Kawakatu \& Kino 2006). 
The evolution of $A_{\rm h}(l_{\rm h})$ could be determined by 
the density profile of ambient medium and the growth rate of cross-section of 
cocoons (e.g., Begelman \& Cioffi 1989; Loken et al. 1992; KK05). 
However, it is difficult to obtain $A_{\rm h}(l_{\rm h})$ analytically 
because this is determined by the non-linear effects (Smith et al. 1985; 
Lind et al. 1989; Scheck et al. 2002; Mizuta et al. 2004; Perucho \& Marti 2007). Then, we assume $A_{\rm h}(l_{\rm h})$ as being proportional to $r_{\rm HS}^{2}
(l_{\rm h})$ 
as follows; 
\begin{equation}
A_{\rm h}(l_{\rm h})\equiv f\pi r_{\rm HS}^{2}(l_{\rm h}), 
\end{equation}
where $r_{\rm HS}(l_{\rm h})$ is the hot spot size. Although the parameter $f$
($ > 1$) is determined by the non-linear effects (e.g., the vortex and oblique shock) at the hot spots (e.g., Mizuta et al. 2004), the 2D relativistic numerical simulations  show $f=const.$ (e.g., Scheck et al. 2002; Perucho \& Marti 2007). Thus, we here suppose $f=const.$.

For the ambient density profile $\rho_{\rm a}(l_{\rm h})$, we assume a double 
power-law distribution obtained from X-ray observations (e.g., Trinchieri et al. 1986; Mathews \& Brighenti 2003; Allen et al. 2006; Fukazawa et al. 2006) as 
\begin{equation}
\rho_{\rm a}(l_{\rm h})=\left \{
\begin{array}{l}
  \bar{\rho}_{\rm a}(l_{\rm h}/l_{\rm c})^{-\alpha_{\rm inner}} \, \,\,\, 
  {\rm for}\,\,\, l_{\rm h} < l_{\rm c}, \\ \\
  \bar{\rho}_{\rm a}(l_{\rm h}/l_{\rm c})^{-\alpha_{\rm outer}} \, \,\,\, 
  {\rm for}\,\,\, l_{\rm h} > l_{\rm c}, 
\end{array}
\right .
\end{equation}
%
where $\bar{\rho}_{\rm a}$ is the mass density of the ambient 
matter at $l_{\rm c}$ where the slope index of $\rho_{\rm a}(l_{\rm h})$ 
changes and $\alpha \geq 0$ is the slope index of $\rho_{\rm a}(l_{\rm h})$. 
Here $\bar{\rho}_{\rm a}=\bar{n}_{\rm a}m_{\rm p}$ where $m_{\rm p}$ 
is the proton mass. Based on the observed temperature profile 
of elliptical galaxies from X-ray observations 
(e.g., Allen et al. 2001; Fukazawa et al. 2006), 
we assume the ambient temperature profile $T_{\rm a}(l_{\rm h})$ 
as follows;
\begin{equation}
T_{\rm a}(l_{\rm h})=\left \{
\begin{array}{l}
  T_{\rm a,c}\, \,\,\, 
  {\rm for}\,\,\, l_{\rm h} < l_{\rm c}, \\ \\
  T_{\rm a,c}(l_{\rm h}/l_{\rm c})^{\beta} \, \,\,\, 
  {\rm for}\,\,\, l_{\rm c} < l_{\rm h} < l_{\rm g}, \\ \\
  T_{\rm a,g}\, \,\,\, 
  {\rm for}\,\,\, l_{\rm h} > l_{\rm g},
\end{array}
\right .
\end{equation}
where $T_{\rm a,g}=T_{\rm a, c}(l_{\rm g}/l_{\rm c})^{\beta}$ with $\beta \geq 
0$ and $l_{\rm g}$ is the distance where $T_{\rm a}(l_{\rm h})=T_{\rm a, g}$. 
%
%

From the radio observations (Jeyakumar \& Saikia 2000; KNK08), 
the evolution of $r_{\rm HS}(l_{\rm h})$ can be assumed to be a broken power-law distribution as 
\begin{equation}
r_{\rm HS}(l_{\rm h})=\left \{
\begin{array}{l}
  \bar{r}_{\rm HS}(l_{\rm h}/l_{\rm c})^{\gamma_{\rm inner}}\, \,\,\, 
  {\rm for}\,\,\, l_{\rm h} < l_{\rm c}, \\ \\
  \bar{r}_{\rm HS}(l_{\rm h}/l_{\rm c})^{\gamma_{\rm outer}}\, \,\,\, 
  {\rm for}\,\,\, l_{\rm h} > l_{\rm c}, 
\end{array}
\right .
\end{equation}
where $\bar{r}_{\rm HS}$ is the hot spot radius at $l_{\rm c}$. 
%

\section{On the origin of FRI-FRI\hspace{-.1em}I dichotomy}
Based on simple treatments for the dynamical evolution of radio sources, 
we will derive the condition in which initially supersonic lobes 
turn into subsonic lobes. We suppose that the growth of 
radio sources changes drastically when the advance speed along the jet axis 
equals to the sound speed of the ambient medium, $c_{\rm s}(l_{\rm h})$ 
(e.g., Gopal-Krishna \& Wiita 1987; Gopal-Krishna \& Wiita 1991). 
This criterion is simply described as 
\begin{equation}
v_{\rm HS}(l_{\rm h})= c_{\rm s}(l_{\rm h}). 
\end{equation}
Here $c_{\rm s}(l_{\rm h})=(5kT_{\rm a}(l_{\rm h})/3m_{\rm p})^{1/2}$ 
where $k$ is the Boltzman constant.
Using eqs. (2) and (6), the eq. (1) can be expressed 
as follows; 
\begin{eqnarray}
L_{\rm j}/c&=&f\pi r_{\rm HS}^{2}(l_{\rm h})\rho_{\rm a}(l_{\rm h})
c_{\rm s}^{2}(l_{\rm h}).
\end{eqnarray}
By substituting eqs. (3), (4) and (5) into eq. (7), 
the critical line between the supersonic lobes and subsonic ones 
is expressed as a function of $l_{\rm h}$ as follows; 
\begin{equation}
\left(\frac{L_{\rm j}}{\bar{n}_{\rm a}} \right)_{\rm crit}= 
fcm_{\rm p}\bar{r}^{2}_{\rm HS}
\left \{
\begin{array}{l}
c^{2}_{\rm s, c}(l_{\rm h}/l_{\rm c})^{\,\,(2\gamma-\alpha)_{\rm inner}}\, \,\,\,
{\rm for}\,\,\, l_{\rm h} < l_{\rm c}, \\ \\
c^{2}_{\rm s, c}(l_{\rm h}/l_{\rm c})^{\,\,(2\gamma-\alpha)_{\rm outer}+\beta}\, \,\,\,
{\rm for}\,\,\, l_{\rm c} < l_{\rm h} < l_{\rm g}, \\ \\
c^{2}_{\rm s, g}(l_{\rm h}/l_{\rm c})^
{\,\,(2\gamma-\alpha)_{\rm outer}}\, \,\,\,
{\rm for}\,\,\, l_{\rm h} > l_{\rm g}, 
\end{array}
\right .
\end{equation}
where $c_{\rm s,i}(l_{\rm h})=(5kT_{\rm a,i}(l_{\rm h})/3m_{\rm p})^{1/2}$ 
with i=``c'' or ``g'', $(2\gamma-\alpha)_{\rm inner}\equiv 2\gamma_{\rm inner}-
\alpha_{\rm inner}$ and $(2\gamma-\alpha)_{\rm outer}\equiv 2\gamma_{\rm outer}
-\alpha_{\rm outer}$. 

\vspace{2mm}
\epsfxsize=8cm 
\epsfbox{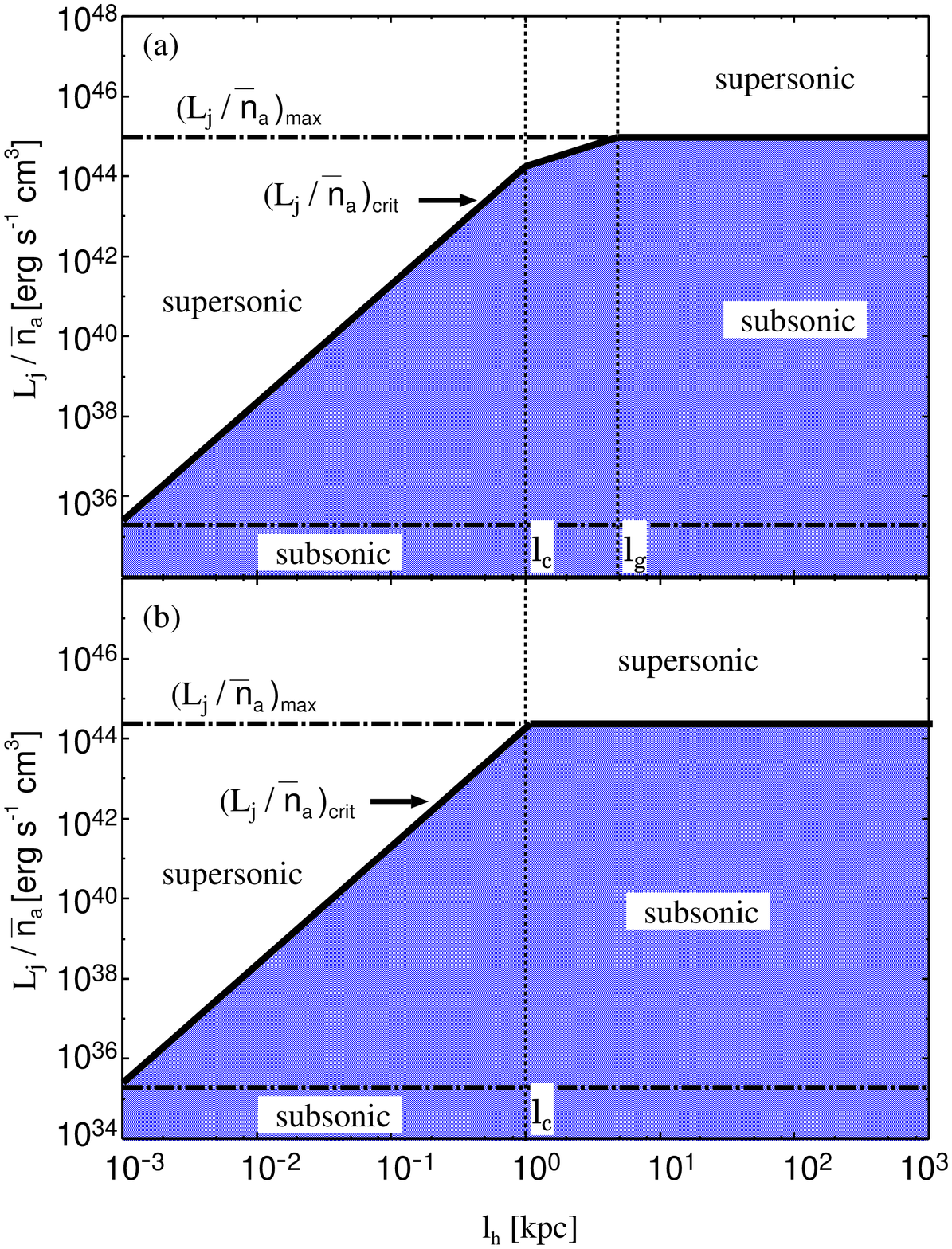}
\figcaption
{
(a) The ratio of the jet's kinetic power and the constant number density of the ambient medium in over the core radius ($l_{\rm c}=$1 kpc), $L_{\rm j}/\bar{n}_{\rm a}$, against the distance from the core, $l_{\rm h}$ for $f=10$, $(2\gamma-\alpha)_{\rm inner}=3$, $(2\gamma-\alpha)_{\rm outer}=0$ and $l_{\rm c}=1\,{\rm kpc}$. 
We assume $T_{\rm a,c}=2\times 10^{6}\,{\rm K}$ and $T_{\rm a,g}=1\times 
10^{7}\,{\rm K}$. 
The solid line $[(L_{\rm j}/\bar{n}_{\rm a})_{\rm crit}]$ determined by eq. (10) shows the dividing line between the supersonic lobes and the subsonic ones (the shaded region) 
for $\beta=1$ where $\beta$ is the slope index of $T_{\rm a}(l_{\rm h})$ for 
$l_{\rm c} < l_{\rm h} < l_{\rm g}$. The two dot-dashed lines divide into three regions which correspond to the evolutionary 
sequences of variously sized radio galaxies (see $\S4$). 
(b) Same as (a) but for $\beta=0$ and $T_{\rm a, c}=T_{\rm a, g}
=2\times 10^{6}\,{\rm K}$. 
}
\vspace{2mm}

To determine the critical line, the slope index of $(L_{\rm j}/\bar{n}_{\rm a})_{\rm crit}$ is important. Since the slope index depends on the values of 
$\alpha$ and $\gamma$, we can classify the following three cases, i.e.,  
(i) $(2\gamma-\alpha)_{\rm inner} \geq 0$ and $(2\gamma-\alpha)_{\rm outer} 
\leq 0$, 
(ii) $(2\gamma-\alpha)_{\rm inner} \geq 0$ and $(2\gamma-\alpha)_{\rm outer} > 0$ and 
(iii) $(2\gamma-\alpha)_{\rm inner} < 0$ and any $(2\gamma-\alpha)_{\rm outer}$. 
For the slope of ambient medium $\alpha$, it is possible to constrain as 
$0\leq \alpha_{\rm inner}\leq \alpha_{\rm outer}=1-2$ from X-ray observations (e.g., Trinchieri et al. 1986; Mathews \& Brighenti 2003; Allen et al. 2006; Fukazawa et al. 2006). As for $r_{\rm HS}(l_{\rm h})$, KNK08 found $\gamma_{\rm inner}=1-1.5$ and $\gamma_{\rm outer}=0.3-0.5$ by using about 120 radio sources. 
From these values of $\alpha$ and $\gamma$, the allowed ranges are given by 
\begin{equation}
0\leq (2\gamma-\alpha)_{\rm inner}\leq 3 \,\,\,{\rm and}\,\, 
-1.5\leq (2\gamma-\alpha)_{\rm outer}\leq 0. 
\end{equation}
Thus, the cases (ii) and (iii) can be safely ruled out, then hereafter we will consider the case (i). In case (i), there exists the maximal ratio of the jet power and the number density of the ambient medium at $l_{\rm c}$, $(L_{\rm j}/\bar{n}_{\rm a})_{\rm max}$ because of $(2\gamma-\alpha)_
{\rm inner}\geq 0$ and $(2\gamma-\alpha)_{\rm outer} \leq 0$. 
Importantly, the presence of $(L_{\rm j}/\bar{n}_{\rm a})_{\rm max}$ is 
independent of the values of $\alpha$, $\beta$ and $\gamma$ for case (i), 
except for $(2\gamma-\alpha)_{\rm inner}=0$ and $\beta=0$. 
The existence of $(L_{\rm j}/\bar{n}_{\rm a})_{\rm max}$ is physically 
understood in terms of the strong deceleration of radio lobes 
and hot spots in host galaxies due to the growth of the cross-sectional 
area of radio lobes (see KNK08). The independent observations also imply 
the deceleration of the advance speed of hot spots (O'Dea \& Baum 1997; 
Labiano 2008). Note that the maximum of $(L_{\rm j}/\bar{n}_{\rm a})$ 
disappears only for $\alpha_{\rm inner}=\alpha_{\rm outer}=2$, $\beta=0$, 
$\gamma_{\rm inner}=1$ and $\gamma_{\rm outer}\leq 1$ because $(L_{\rm j}/\bar{n}_{\rm a})_{\rm crit}$ is independent of $l_{\rm h}$. 
If this is the case, the division between FRIs and FRI\hspace{-.1em}Is is 
determined by $T_{\rm a}$ and $r_{\rm HS}$ at the distance from nuclei where the $A_{\rm h}$ growth phase starts, i.e., a few pc.

Figure 1 shows the critical lines (the solid lines) for $\beta=1$ [case (a)] 
and for $\beta=0$ [case (b)], 
assuming $f=10$, $(2\gamma-\alpha)_{\rm inner}=3$, $(2\gamma-\alpha)_
{\rm outer}=0$ and $l_{\rm c}=1\,{\rm kpc}$. 
Concerning $\beta$, Fukazawa et al. (2006) found that the temperature slightly increases with $l_{\rm h}$, i.e., $0 \leq \beta \leq 1$ ($l_{\rm c} <l_{\rm h}<l_{\rm g}$), for X-ray luminous elliptical galaxies, while for X-ray faint 
galaxies shows a flat temperature profile against $l_{\rm h}$, i.e., $\beta=0$. For $T_{\rm a}$ (e.g., Allen et al. 2001; Fukazawa et al. 2006), we assume $T_{\rm a,c}=2\times 10^{6}\,{\rm K}$ and $T_{\rm a,g}=1\times 10^{7}\,{\rm K}$ for case (a) and $T_{\rm a,c}=T_{\rm a,g}=2\times 10^{6}\,{\rm K}$ 
for case (b). For case (a), $l_{\rm g}=5\,{\rm kpc}$ because of $l_{\rm g}/l_{\rm c}=T_{\rm a, g}/T_{\rm a, c}$. 
As we mentioned, we see that the maximum of $L_{\rm j}/\bar{n}_{\rm a}$ 
dividing between the supersonic lobes and the subsonic lobes (the shaded region in Figure 1) appears at $l_{\rm c}$. 
From the eq. (8), the maximal ratio of the jet power and the number density at $l_{\rm c}$ can be described as $(L_{\rm j}/\bar{n}
_{\rm a})_{\rm max}=fcm_{\rm p}\pi\bar{r}^{2}_{\rm HS}c^{2}_{\rm s,i}$ 
where i=``g'' for case (a) and i=``c'' for case (b). 
The $(L_{\rm j}/\bar{n}_{\rm a})_{\rm max}$ is expressed as 
\begin{equation}
\left(\frac{L_{\rm j}}{\bar{n}_{\rm a}} \right)_{\rm max}=1\times 10^{45}\left(\frac{f}{10}\right)
\left(\frac{c_{\rm s, i}}{10^{-3}c}\right)^{2}
\left(\frac{\bar{r}_{\rm HS}}{0.3\,{\rm kpc}}\right)^{2}
\,{\rm erg}\,{\rm s}^{-1}\,{\rm cm}^{3}, 
\end{equation} 
where $c_{\rm s, i}=1\times 10^{-3}c$ corresponds to $T_{\rm a, i}=1\times 10^{7}\,{\rm K}$ and $\bar{r}_{\rm HS}=0.3\,{\rm kpc}$ which is derived from Figure 1 in KNK08. Note that $(L_{\rm j}/\bar{n}_{\rm a})_{\rm max}$ for case (b) is 
factor 5 smaller than that for case (a) because of $(L_{\rm j}/\bar{n}_{\rm a})_{\rm max}\propto T_{\rm a}$. 
The predicted $(L_{\rm j}/\bar{n}_{\rm a})_{\rm max}$ in eq. (10) 
is about three orders of magnitude higher than the ratio below which the entrainment can work efficiently, i.e., $L_{\rm j}/\bar{n}_{\rm a} \leq 10^{42}\,{\rm erg}\,{\rm s}^{-1}\,{\rm cm}^{3}$ (De Young 1993a, b; hereafter DY93a, b). 
Thus, by comparing with the future measurements of $L_{\rm j}$ and 
$\bar{n}_{\rm a}$ for the variously sized radio sources, we will be 
able to reveal which deceleration processes, i.e., the jet's head growth 
or entrainment, can divide FRIs and FRI\hspace{-.1em}Is.

Our results suggests that the supersonic lobes can be maintained up to 
$\sim 1$ Mpc when $L_{\rm j}/\bar{n}_{\rm a}$ is larger than its 
maximal value. On the other hand, in the case of below 
$(L_{\rm j}/\bar{n}_{\rm a})_{\rm max}$, the lobes are initially supersonic 
but the supersonic lobes decelerate and then turn into subsonic lobes 
in the core radius of the host galaxy, or the lobes are initially subsonic. 
The division between FRIs and FRI\hspace{-.1em}Is can be determined 
by $(L_{\rm j}/\bar{n}_{\rm a})_{\rm max}$, 
because the radio lobes of FRIs and FRI\hspace{-.1em}Is 
can be subsonic and supersonic, respectively. 
This indicates that FRIs favor the higher $\bar{n}_{\rm a}$ than 
FRI\hspace{-.1em}Is at given $L_{\rm j}$. 
This is consistent with the observational results 
showing that FRIs are located at the centers of clusters of galaxies 
while FRI\hspace{-.1em}Is are discovered in the fields or 
at the edge of clusters of galaxies (e.g., Prestage \& Peacock 1988; 
Miller et al. 2002). 
The division in the Owen-White diagram has been shown to be 
much less clear based on the latest Sloan Digital Sky Surveys sample 
(Best et al. 2009). 
This may indicate that radio luminosity and optical magnitude of the 
host galaxy are not the fundamental physical quantities for the demarcation 
between FRIs and FRI\hspace{-.1em}Is. 
In order to judge whether our prediction is reasonable, it will be essential 
to compare with the observed $L_{\rm j}$ and $\bar{n}_{\rm a}$ for 
the variously sized radio galaxies (i.e., CSOs, MSOs, FRIs and FRI\hspace{-.1em}Is). 

Lastly, we discuss how $(L_{\rm j}/\bar{n}_{\rm a})_{\rm max}$ depends on 
the different choice of parameters, i.e., $f$, $T_{\rm a}$ and 
$\bar{r}_{\rm HS}$. 
As for parameter $f$, $A_{\rm h}$ must be smaller than $\pi l_{\rm h}^{2}$ 
because of the elongated morphology of radio lobes. 
According to the observed relation between the hot spot size, 
$r_{\rm HS}$ and the projected linear size, $l_{\rm h}$ relation 
(see Figure 1 in KNK08) $f$ can be $10 < f < 10^{2}$. 
For four FRI\hspace{-.1em}Is, by directly comparing 
$\pi r_{\rm HS}^{2}$ with $A_{\rm h}$, the range of $f$ can be obtained 
as $ 10 < f < 10^{2}$ (e.g., Ito et al. 2008; see Carilli et al. 1991). 
Thus, we here consider a wide range of $f$, i.e., $10 < f < 10^{2}$. 
As $c_{\rm s}$ becomes larger, $(L_{\rm j}/\bar{n}_{\rm a})_{\rm max}$ 
increases at fixed $\bar{r}_{\rm HS}$ and vice versa (see eq. (10)). 
In order to satisfy $v_{\rm HS}\geq c_{\rm s}$, 
the larger $(L_{\rm j}/\bar{n}_{\rm a})_{\rm max}$ is required  
as $c_{\rm s}$ increases because of $v_{\rm HS} \propto L_{\rm j}/
\bar{n}_{\rm a}$ (see eq. (1)). 
According to the current observations, the ranges of temperature could be 
narrow, i.e., $2\times 10^{6}\,{\rm K}\leq T_{\rm a,c} \leq 1\times 10^{7}\,{\rm K}$ and $1\times 10^{7}\,{\rm K}\leq T_{\rm a,g} \leq 2\times 10^{7}\,{\rm K}$
(e.g., Allen et al. 2001; Fukazawa et al. 2006). 
For the hot spot radius at $l_{\rm c}$, the range of $\bar{r}_{\rm HS}$ 
could be $0.3\,{\rm kpc} \leq \bar{r}_{\rm HS} \leq 1\,{\rm kpc}$ (see Figure 1 in KNK08) because of $1\,{\rm kpc} \leq l_{\rm c} \leq 10\,{\rm kpc}$ 
(e.g., Fukazawa et al. 2006). Considering all uncertainties, the allowed range 
of $(L_{\rm j}/\bar{n}_{\rm a})_{\rm max}$ can be $\approx 10^{44}-10^{47}
\,{\rm erg}\,{\rm s}^{-1}\,{\rm cm}^{3}$, which is more than 2 orders of magnitude higher than the ratio below which the entrainment would be important 
(e.g., DY93a, b). 
To summarize, it will be essential to measure $\rho_{\rm a}(l_{\rm h})$, 
$T_{\rm a}(l_{\rm h})$ and $f$ more accurately, in order to determine 
a critical line dividing between FRIs and FRI\hspace{-.1em}Is.

\section{Predictions: Evolutionary tracks of radio galaxies}
In order to evolve into the large-scale [$l_{\rm h}> l_{\rm c}(\leq 10\,{\rm kpc})$] supersonic lobes, it is necessary to hold the condition 
which $L_{\rm j}/\bar{n}_{\rm a}$ is larger than 
$(L_{\rm j}/\bar{n}_{\rm a})_{\rm max}$. 
Moreover, the dividing line between the subsonic and supersonic, 
$(L_{\rm j}/\bar{n}_{\rm a})_{\rm crit}$ depends on $l_{\rm h}$. 
Thus, the evolutionary path of radio sources can be divided by 
$(L_{\rm j}/\bar{n}_{\rm a})_{\rm crit}$ (see eq. (8)). 
Observationally, several authors have discovered young and compact radio 
sources such as compact symmetric objects (CSOs; $l_{\rm h} <$ 1kpc) and 
medium-size symmetric objects (MSOs; $l_{\rm h}=1-10$ kpc) (e.g., Wilkinson et al. 1994; Fanti et al. 1995; Readhead et al. 1996). 

On the basis of our findings (see Figure 1), we can predict the fate 
of compact and young radio galaxies as follows;  
\begin{description}
\item [(i)]
If $L_{\rm j}/\bar{n}_{\rm a} > (L_{\rm j}/\bar{n}_{\rm a})_{\rm max}
=10^{44-45}\,{\rm erg}\,{\rm s}^{-1}\,{\rm cm}^{3}$, 
the evolutionary sequence appears as 
CSOs $\to$ MSOs $\to$ FRI\hspace{-.1em}Is. 

\item [(ii)]
When $L_{\rm j}/\bar{n}_{\rm a} < (L_{\rm j}/\bar{n}_{\rm a})_{\rm max}
=10^{44-45}\,{\rm erg}\,{\rm s}^{-1}\,{\rm cm}^{3}$, 
the evolutionary track is as follows; 
CSOs $\to$ distorted MSOs $\to$ FRIs. 

\item [(iii)]
If $(L_{\rm j}/\bar{n}_{\rm a}) < 10^{36}\,{\rm erg}\,{\rm s}^{-1}\,{\rm cm}^{3}$ at $l_{\rm h}=1\,{\rm pc}$, the evolutionary track appears as the FRI-like 
CSOs $\to$ distorted MSOs $\to$ FRIs. 
\end{description}
For case (iii), the distorted MSOs might correspond to $\sim 1\,{\rm kpc}$ 
low power compact (LPC) radio sources (e.g., Kunert-Bajraszewska et al. 2005; 
Giroletti et al. 2005) because the radio morphology of LPCs 
tends to be irregular and some LPCs show FRI-like morphology (e.g., Giroletti et al. 2005). 
Note that the deceleration of supersonic lobes via the entrainment 
may be also important for cases (ii) and (iii).

We briefly comment on the number of radio sources with supersonic lobes 
per bin of projected size, $N(l_{\rm h})$. As seen in Figure 1, 
the region of the supersonic lobes allowed is larger as the size of 
the AGN jets is smaller. 
This might imply that the number of CSOs having supersonic lobes 
is larger than that of MSOs. However, in order to predict $N(l_{\rm h})$ 
we need to consider the luminosity evolution of radio sources, 
the distribution of $L_{\rm j}$ and $\bar{n}_{\rm a}$. 
This is left as our future work.

\section{Summary}
We examine the origin of `FRI/FRI\hspace{-.1em}I dichotomy' 
by considering the deceleration of expanding radio lobes. 
We explored the condition of a supersonic lobe turnng into a subsonic one, comparing the observed $r_{\rm HS}-l_{\rm h}$ relation with the 
ram-pressure confinement along the jet axis. 
We then found that the dividing line between the supersonic lobes 
and subsonic ones is determined by the single parameter, i.e., the ratio of the jet power ($L_{\rm j}$) and the number density of the ambient medium at the core radius of hosts ($\bar{n}_{\rm a}$). 
Importantly, there exists the maximal ratio of $(L_{\rm j}/\bar{n}_{\rm a})$ and its value is $(L_{\rm j}/\bar{n}_{\rm a})_{\rm max}\approx 10^{44-47}\,{\rm erg}\, {\rm s}^{-1}\,{\rm cm}^{3}$, taking account of considerable uncertainties. 
This is more than 2 orders of magnitude higher than 
the ratio below which the entrainment would be important 
(e.g., DY93a, b). 
Thus, it will be able to test whether 
the predicted maximal ratio $(L_{\rm j}/\bar{n}_{\rm a})_{\rm max}$ 
divides the FRIs from the FRI\hspace{-.1em}Is, by comparing with the future 
observation of $L_{\rm j}$ and $\bar{n}_{\rm a}$ for the variously sized radio 
sources. 

\acknowledgments 
We thank the anonymous referee for several comments helpful 
toward improving our work. 
NK is financially supported 
by the Japan Society for the Promotion of Science (JSPS) through 
the JSPS Research Fellowship for Young Scientists.

\end{document}